\documentclass[12pt]{article}
\usepackage{epsf}
\def\be{\begin{equation}}
\def\ee{\end{equation}}

\begin{document}
\begin{titlepage}
\begin{center}
{\Large \bf William I. Fine Theoretical Physics Institute \\
University of Minnesota \\}
\end{center}
\vspace{0.2in}
\begin{flushright}
FTPI-MINN-11/13 \\
UMN-TH-3003/11 \\
May 2011 \\
\end{flushright}
\vspace{0.3in}
\begin{center}
{\Large \bf Radiative transitions from $\Upsilon(5S)$ to molecular bottomonium
\\}
\vspace{0.2in}
{\bf M.B. Voloshin  \\ }
William I. Fine Theoretical Physics Institute, University of
Minnesota,\\ Minneapolis, MN 55455, USA \\
and \\
Institute of Theoretical and Experimental Physics, Moscow, 117218, Russia
\\[0.2in]

\end{center}

\vspace{0.2in}

\begin{abstract}
The heavy quark spin symmetry implies that in addition to the recently observed $Z(10610)$ and $Z(10650)$ molecular resonances with $I^G=1^+$ there should exist two or four molecular bottomonium-like states with $I^G=1^-$. Properties of these $G$-odd states are considered, including their production in the radiative transitions from $\Upsilon(5S)$, by applying the same symmetry to the $\Upsilon(5S)$ resonance and the transition amplitudes. The considered radiative processes can provide a realistic option for observing the yet hypothetical states.

\end{abstract}
\end{titlepage}

The masses and the decay properties of the recently discovered~\cite{belle} isotriplet resonances $Z_b(10610)$ ($Z_b$) and $Z_b(10650)$ ($Z_b'$) at the $B^* \bar B$ and $B^* \bar B^*$ thresholds strongly suggest that these are  `molecular' type~\cite{bgmmv} threshold singularities in the $S$-wave channels for the corresponding heavy meson-antimeson pair. The properties of such threshold resonances are related by the heavy quark spin symmetry (HQSS) similarly to the relations between the properties of the $B$ and $B^*$ mesons. In particular it has been argued~\cite{bgmmv} that the existence of the observed `twin' resonances $Z_b$ and $Z_b'$ resonances and the heavy quark spin symmetry imply an existence of a larger family of `molecular' peaks at the thresholds of the $B$ and $B^*$ meson-antimeson pairs, which family should consist of at least four, but more likely of six isovector resonances. The observed $Z_b$ and $Z_b'$ states have quantum numbers $I^G(J^P)=1^+(1^+)$. The rest of the isovector states in the family have negative $G$ parity and are denoted here as $W_{bJ}$ with two states having $J=0$ ($W_{b0}$ and $W_{b0}'$) and one each with $J=1$ and $J=2$: $W_{b1}$ and $W_{b2}$. The $W_{bJ}$ states cannot be produced in  single-pion transitions from $\Upsilon(5S)$, but can be produced in hadronic transitions with emission of $\rho$ meson from higher $\Upsilon$-like bottomonium states with mass above approximately 11.4\,GeV. At present however no data are available on such bottomonium states, and we have to consider other possibilities for a study of the expected new molecular resonances. In this paper I discuss the production of the isotopically (and electrically) neutral components of the expected isovector multiplets in radiative transitions from the $\Upsilon(5S)$ resonance: $\Upsilon(5S) \to W_{bJ} \, \gamma$. Specifically, will be considered the relations, following from HQSS, between the rates of such transitions to all the $W_{bJ}$ resonances. Although a prediction of the absolute rate is currently quite uncertain and is limited to the general expectation that $\Gamma[\Upsilon(5S) \to W_{bJ} \, \gamma]/\Gamma[\Upsilon(5S) \to Z_b \, \pi] \sim \alpha$, an understanding of the relative rates could serve as a guidance in the searches for the expected molecular resonances.


The relations between the discussed resonances arise due to the known suppression by the inverse of the $b$ quark mass of the strong interaction depending on the spin of the $b$ quark or antiquark. In particular in the limit, where this interaction is completely turned off the spin variables of the heavy quark and antiquark are purely `classificational' in the sense that they define the quantum numbers of the states containing the $b \bar b$ pair, but the dynamics proceeds as if the heavy quark had no spin at all. In particular in this `spinless $b$' (SLB) limit the $B$ mesons behave as heavy particles with spin $1/2$, and an $S$-wave $B \bar B$ pair can be in a state with total angular momentum zero or one, and  these states are denoted here as $0^-_{SLB}$ and $1^-_{SLB}$ respectively, where the superscript indicates the parity. Clearly, in the SLB limit the dynamical properties, most importantly those arising from the interaction of the light components of the $B$ mesons, are determined only by these quantum numbers, so that there are only two independent channels for the $B \bar B$ meson pairs at a fixed total isospin.  In particular, a threshold singularity in one or both channels gives rise to threshold `molecular' resonances for the physical meson-antimeson states. It can be seen~\cite{bgmmv} that for the case where the singularity at the threshold (a bound or a virtual state as appropriate for an $S$-wave scattering) exists only in the isovector $0^-_{SLB}$   channel, the family of the physical resonances consists of two pairs of `twin' peaks: the $Z_b$ and $Z_b'$ resonances, and an additional pair of `twin' resonances with the quantum numbers $I^G(J^P)=1^-(0^+)$:  $W_{b0}$ and $W_{b0}'$, at respectively the $B \bar B$ and the $B^* \bar B^*$ thresholds. If, however, a threshold singularity is in the $1^-_{SLB}$ channel, or in both channels, then, besides those four resonances, also arise two additional states: $W_{b1}$ with $I^G(J^P)=1^-(1^+)$ near the $B^* \bar B$ threshold and $W_{b2}$ with $I^G(J^P)=1^-(2^+)$ near the $B^* \bar B^*$ threshold.   The resonances $W_{b1}$ and $W_{b2}$ contain the heavy $b \bar b$ quark pair in a pure spin-1 (ortho-) state, while each of the pairs of `twin' resonances contains a mixture of the ortho- and para- (spin-0) states of the heavy quark pair. The $Z_b$ and $Z_b'$ resonances are two maximal ($45^\circ$) mixtures, while in the $W_{b0}$ and $W_{b1}$ resonances the ortho - para mixing is with a $30^\circ$ angle. Therefore the HQSS requires that the $W_{b1}$ and $W_{b2}$ have an unsuppressed coupling to channels with ortho-bottomonium, while the $Z_b\, (Z_b')$ and $W_{b0}\, (W_{b0}')$ resonances couple to channels with ortho- as well as with para- bottomonium with the relative coefficients in these couplings determined by the coefficients in the ortho - para mixing within these resonances.

It should be mentioned that if only one of the possible SLB channels possesses a near-threshold singularity then it should be the $0^-_{SLB}$ rather than $1^-_{SLB}$, since it is known~\cite{wg} from the general properties of the QCD that the energy of the ground state in a pseudoscalar flavor-non-singlet channel is not larger than that in the vector channel.

The described picture of the family of the threshold resonances can be established by considering the composition of the SLB spin states $0^-_{SLB}$ and $1^-_{SLB}$ with the spin states of the $b \bar b$ pair with the total spin $0^-_H$ and $1^-_H$ in terms of the physical meson-antimeson pairs containing a $B$ or $B^*$ meson and an antimeson. One can find the appropriate compositions with fixed overall quantum numbers by explicitly reconstructing the eigenstates of the spin-dependent Hamiltonian $H_s$ that lifts the SLB degeneracy between those physical states. The Hamiltonian can be written in therms of the spin operators $\vec s_b$ ($\vec s_{\bar b}$) for the $b$ ($\bar b$) quark and $\vec s_q$ ($\vec s_{\bar q}$) describing  the $B$ ($\bar B$) mesons in the SLB limit (`the spin of the light (anti)quark'):
\be
H_s= \mu \, (\vec s_b \cdot \vec s_{\bar q}) + \mu \, (\vec s_{\bar b} \cdot \vec s_{q})= {\mu \over 2} \, (\vec S_H \cdot \vec S_{SLB}) - {\mu \over 2} \, (\vec \Delta_H \cdot \vec \Delta_{SLB})~,
\label{hs}
\ee
where $\vec S_H = \vec s_b + \vec s_{\bar b}$, $\vec S_{SLB}= \vec s_q+ \vec s_{\bar q}$,  $\vec \Delta_H = \vec s_b - \vec s_{\bar b}$ and $\vec \Delta_{SLB}= \vec s_q - \vec s_{\bar q}$. The first expression in Eq.(\ref{hs}) is the standard phenomenological Hamiltonian for describing the masses of the $B$ and $B^*$ mesons: $M(B)=\bar M - 3 \, \mu/4$, $M(B^*)=\bar M + \mu/4$, with $\bar M$ being the (common) mass of the $B$ and $B^*$ mesons in the SLB limit, so that $\mu= M(B^*)-M(B) \approx 46$\,MeV. The latter form of the expression in Eq.(\ref{hs}) is convenient for considering the states of the meson-antimeson pairs in terms of the total spin in the $SLB$ limit and of the total spin of the $b \bar b$ pair. The convenience of the latter form of presenting the Hamiltonian $H_s$ arises from the fact that the product $(\vec S_H \cdot \vec S_{SLB})$ depends only on the overall total spin of the state $|\vec J|=|\vec S_H + \vec S_{SLB}|$, while the operator $\vec \Delta$ has only non-diagonal matrix elements between the spin-singlet and spin-triplet states. Considering non interacting meson-antimeson pairs, one can use the Hamiltonian $H_s$ to find the eigenstates:
\be
1^-(2^+):~~ \left. \left ( 1^-_H \otimes 1^-_{SLB} \right ) \right |_{J=2}~, ~~~ {1 \over 2} \, \mu~; 
\label{2p}
\ee
\be
1^-(1^+):~~ \left. \left ( 1^-_H \otimes 1^-_{SLB} \right ) \right |_{J=1}~, ~~~ - {1 \over 2} \, \mu~;
\label{1p}
\ee
\be
1^-(0^+):~~ {\sqrt{3} \over 2} \, \left ( 0^-_H \otimes 0^-_{SLB} \right ) + {1 \over 2} \,\left. \left ( 1^-_H \otimes 1^-_{SLB} \right ) \right |_{J=0}~, ~~~  {1 \over 2} \, \mu~;
\label{0p2}
\ee
\be
1^-(0^+):~~ {1 \over 2} \, \left ( 0^-_H \otimes 0^-_{SLB} \right ) - {\sqrt{3} \over 2} \,\left. \left ( 1^-_H \otimes 1^-_{SLB} \right ) \right |_{J=0}~, ~~~  -{3 \over 2} \, \mu~;
\label{0p1}
\ee
\be
1^+(1^-):~~ {1 \over \sqrt{2}} \, \left ( 0^-_H \otimes 1^-_{SLB} \right ) - {1 \over \sqrt{2}} \, \left ( 1^-_H \otimes 0^-_{SLB} \right ) ~, ~~~  {1 \over 2} \, \mu~;
\label{zbp}
\ee
\be
1^+(1^-):~~ {1 \over \sqrt{2}} \, \left ( 0^-_H \otimes 1^-_{SLB} \right ) + {1 \over \sqrt{2}} \, \left ( 1^-_H \otimes 0^-_{SLB} \right )~, ~~~  -{1 \over 2} \, \mu~.
\label{zb}
\ee
In these formulas the first expression indicates the quantum numbers $I^G(J^P)$, the second describes the composition in terms of the heavy and SLB spin states, and the third one gives the energy of the state relative to the SLB $B \bar B$ threshold $2 \bar M$. Considering these energy shifts, one readily identifies the states (\ref{2p}), (\ref{0p2}) and (\ref{zbp}) as being at the $B^* \bar B^*$ threshold, the states (\ref{1p}) and (\ref{zb}) at the $B^* \bar B$ threshold and, finally, the state (\ref{0p1})  at the $B \bar B$ threshold.

The major effect of the interaction depending on the spin of the heavy quark is the splitting of the (otherwise degenerate) thresholds for the scattering channels described by Eqs.~(\ref{2p}) - (\ref{zb}). The actual interaction between mesons takes place in a finite volume, where the effect of the forces related to the spin of the heavy quark is small in comparison with the interaction giving rise to near threshold singularities in either one or both the $0^-_{SLB}$ and $1^-_{SLB}$ channels. Thus one comes to the conclusion that the expected family of the near threshold molecular resonances is as shown in Fig.1. The existence of the $J^P=0^+$ states $W_{b0}$ and $W_{b0}'$ follows from the existence of the $Z_b(Z_b')$ resonances, while the existence of the $W_{b1}$ and $W_{b2}$ is contingent on the presence of a near threshold singularity in the $1^-_{SLB}$ channel. It can be also noted that the $W_{b1}$ state is a pure isovector bottomonium-like analog of the charmonium-like resonance X(3872), which is a pure $(1^-_H \otimes 1^-_{SLC})$ state~\cite{mv}.

\begin{figure}[ht]
\begin{center}
 \leavevmode
    \epsfxsize=11cm
    \epsfbox{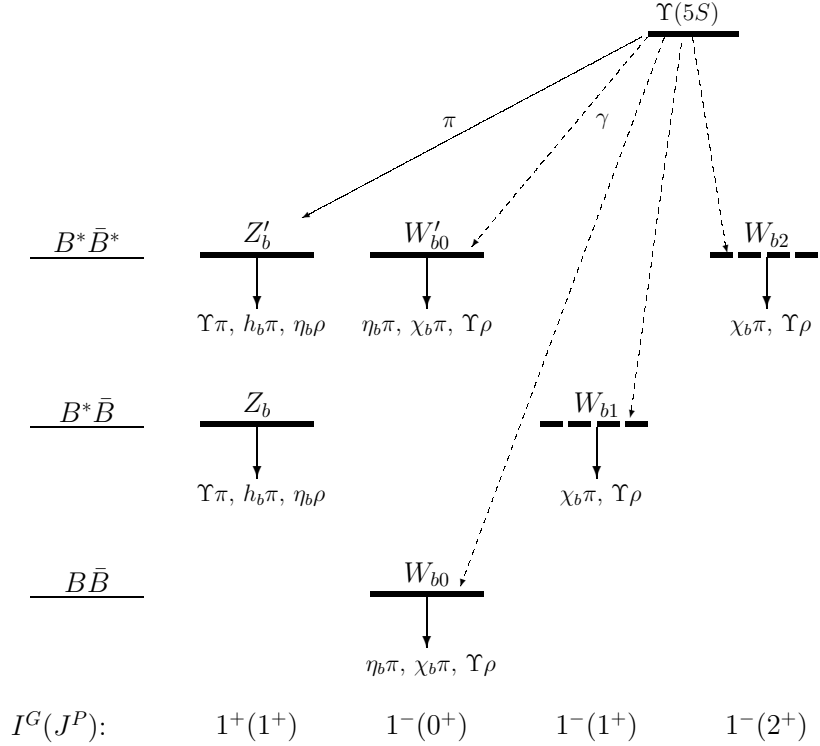}
    \caption{The expected family of six isotriplet resonances at the $B \bar B$, $B^* \bar B$ and $B^* \bar B^*$ thresholds and their likely decay modes to bottomonium and a light meson. The excited bottomonium states can be present in the decays instead of the shown lower states ($\eta_b, \, \Upsilon, \, h_b, \, \chi_b$), where kinematically possible. The dashed arrowed lines show the discussed radiative transitions from $\Upsilon(5S)$. (The mass splitting to $\Upsilon(5S)$ is shown not to scale.)}
\end{center}
\end{figure} 

Clearly, the $H \otimes SLB$ spin structure described be Eqs.~(\ref{2p}) - (\ref{zb}) also implies relations between the total widths of the $W_{bJ}$ states:
\be
\Gamma(W_{b2})=\Gamma(W_{b1})={3 \over 2} \, \Gamma(W_{b0}) - {1 \over 2} \, \Gamma(W_{b0}')
\label{gtr}
\ee
as well as relations for
the rates of decays of the resonances to specific channels, e.g.
\be
\Gamma(W_{b0} \to \Upsilon \rho):\Gamma(W_{b0}' \to \Upsilon \rho):\Gamma(W_{b1} \to \Upsilon \rho) : \Gamma(W_{b2} \to \Upsilon \rho) = {3 \over 4} : {1 \over 4} : 1 :1~,
\label{gr}
\ee
with a possible slight modification due to the kinematical difference in the phase space. Such exclusive decays can be used for an experimental identification of the resonances. This paper however concentrates on the radiative transitions in which the $W_{bJ}$ states can be produced. 


The details of the transitions from $\Upsilon(5S)$ crucially depend on its structure in terms of its decomposition into heavy (H) and SLB spin states. 
The resonance $\Upsilon(5S)$ is produced in $e^+e^-$ annihilation by the electromagnetic current, which creates a $b \bar b$ pair in an orhtho state $1^-_H$. In the standard notation, the heavy quark pair is created in either a $^3S_1$ state, or in a $^3D_1$ one. In terms of the $H \otimes SLB$ decomposition the former is $1^-_H \otimes 0^+_{SLB}$ and the latter is $1^-_H \otimes 2^+_{SLB}$, since the angular momentum of the $b \bar b$ pair is relegated to the `SLB' system. However the $D$-wave contribution in the production of the resonance is small inasmuch as the heavy quarks are nonrelativistic at the energy of the $\Upsilon(5S)$ resonance and can be neglected. It is thus reasonable to assume that the spin structure of $\Upsilon(5S)$ in terms of a $H \otimes SLB$ decomposition is dominated by $1^-_H \otimes 0^+_{SLB}$.

To a certain extent the assumed spin structure of the resonance can be tested against the available date on its decays. Namely, the relative yield of the meson-antimeson pairs $B^* \bar B^*$, $B^* \bar B + B \bar B^*$, and $B \bar B$ significantly depends on this structure. In the limit, where the interaction of the heavy quark spin is considered as small for a pure $1^-_H \otimes 0^+_{SLB}$ state the ratio of the yield in these channels is 7:3:1~\cite{drgg} (see also in the review~\cite{mvc}). In the real world there are finite effects due to the spin-dependent interaction both in the decay amplitudes and in the kinematical $P$-wave factors $p^3$ due to the mass splitting between the $B^*$ and $B$ mesons due to the same interaction. If only the phase space factors $p^3$ are taken into account the ratio becomes 4.2:2.4:1. However, given the absence of a full calculation in the first order in the spin effects it may be more reasonable to compare the lowest order theoretical result with the experimental data. Thus the spread between the expected ratio with and without the kinematical factors illustrates the range of current theoretical uncertainty. The fraction for the yield in each of the three meson-antimeson channels at the $\Upsilon(5S)$ resonance as measured by Belle~\cite{belle2} corresponds to $f(B^* \bar B^*)=(37.5^{+2.1}_{-1.9}\pm3.0)\%,~ f(B^* \bar B + B \bar B^*)=(13.7 \pm 1.3 \pm 1.1)\%,~ f(B \bar B) = (5.5^{+1.0}_{-0.9}\pm 0.4)\%$, which reasonably agrees with the 7:3:1 ratio, and given the errors, with the kinematically modified ratio. In either case, the suggested spin structure $1^-_H \otimes 0^+_{SLB}$ of the $\Upsilon(5S)$ resonance appears to not contradict the data.

Another test of the suggested spin structure of $\Upsilon(5S)$ is provided by its decays into the channels $B^* \bar B^* \pi $ $(B^* \bar B + B \bar B^*) \pi$ and $B \bar B \pi$. The energy above the threshold for the heavy meson pair in these processes is small, so that only the lowest possible partial wave amplitude can be retained when considering these decays.  This is in agreement with the observed~\cite{belle2} suppression of the channel $B \bar B \pi$: $f(B \bar B \pi)= (0.0 \pm 1.2 \pm 0.3)\%$, since this process cannot go in the $S$-wave unlike the two other. In the $S$ wave the states of the heavy meson pair in terms of the $H \otimes SLB$ decomposition can be read off the formulas (\ref{zbp}) for $B^* \bar B^*$ and (\ref{zb}) for  $(B^* \bar B + B \bar B^*)$.
The heavy quark spin state is conserved, so that the decays from $\Upsilon(5S)$ proceed only to the $1^-_H \otimes 0^-_{SLB}$ component of the states of these heavy meson pairs. In other words the underlying process can be viewed as factorized into the transition $(1^-_H)_{\Upsilon(5S)} \to (1^-_H)_{\rm final}$ for the heavy spin and $(0^+_{SLB})_{\Upsilon(5S)} \to (0^-_{SLB})_{\rm final} + \pi$ for the rest degrees of freedom. Clearly, since the states (\ref{zbp}) and (\ref{zb}) contain the spin state $1^-_H \otimes 0^-_{SLB}$ with the same amplitude (up to the sign), the ratio of the decay amplitudes of the observed processes can be found as
\be
\left | {A[\Upsilon(5S) \to B^* \bar B^* \, \pi (p_2)] \over A[\Upsilon(5S) \to (B^* \bar B + B \bar B^*) \, \pi (p_1)]} \right | = {E_2 \over E_1}~,
\label{ramp}
\ee
where $p_1$ and $p_2$ are the momenta of the pion in these two decays, and $E_1$ and $E_2$ are the corresponding energies. (The proportionality of the $S$-wave amplitude to the pion energy is dictated by the chiral algebra.) Also in Eq.(\ref{ramp}) it is implied that state $(B^* \bar B + B \bar B^*)$ is the $G=+1$ state of the heavy meson pairs normalized to one. In these processes the kinematical effect of the mass splitting between the $B^*$ and $B$ mesons is considerably enhanced by a very small released kinetic energy: about 75\,MeV in the decay $\Upsilon(5S) \to B^* \bar B^* \, \pi $ and about 120\,MeV in $\Upsilon(5S) \to (B^* \bar B + B \bar B^*) \, \pi $. It thus appears reasonable to take this kinematical effect into account. The estimate for the ratio of the total decay rates further depends on the distribution of the rate over the Dalitz plot. One can estimate the ratio of the relative yield $f(B^* \bar B^* \pi)/ f(B^* \bar B \pi + B \bar B^* \pi) $ as approximately 1/2 if the decay goes dominantly into the lowest invariant mass of the heavy meson pair (i.e. if it is in fact dominated by the $Z_b$ and $Z_b'$ resonances), and as approximately 1/4 if the spectrum of the invariant masses of the heavy pair is given by the phase space. Experimentally~\cite{belle2} the total fractional rates are: $f(B^* \bar B^* \pi) = (1.0 ^{+1.4}_{-1.3} \pm 0.4)\%,~ f(B^* \bar B \pi + B \bar B^* \pi) = (7.3 ^{+2.3}_{-2.1} \pm 0.8)\%$. To the best of my knowledge a Dalitz analysis for these decays in not yet available. Given a visible relative suppression of the channel $B^* \bar B^* \pi$ one can expect, in the suggested here approach, that the Dalitz distribution in these decays should be spread over the physical region, rather than being dominated by the $Z_b$ and $Z_b'$ resonances.


The radiative decays $\Upsilon(5S) \to W_{bJ} \, \gamma$ can be considered in terms of the $H \otimes SLB$ decomposition in the same way as the decays into heavy meson pairs. Indeed, an emission of the photon by the spin of the heavy quarks is negligible, so that in these decays, as before, the $1^-_H$ spin state `goes through'without a change in the orientation of the spin, while the photon emission occurs in the process $(0^+)_{\Upsilon(5S)} \to (1^-_{SLB})_{\rm final} + \gamma$, whose polarization structure is described by only one amplitude in terms of the polarization amplitudes of the $1^-_{SLB}$ state ($\vec \psi$) and of the photon ($\vec a$):
\be
A[(0^+)_{\Upsilon(5S)} \to (1^-_{SLB})_{\rm final} + \gamma] = C \, \omega \, (\vec \psi \cdot \vec a)~,
\label{agam}
\ee
where $\omega$ is the photon energy, and $C$ is a constant, currently unknown, except its  obvious dependence on fine structure constant: $C^2 \propto \alpha$.
Using this structure of the amplitude and the amplitudes of the state $1^-_H \otimes 1^-_{SLB}$ in the $W_{bJ}$ resonances, described by the relations (\ref{2p}) - (\ref{0p1}), one can readily find the ratio of the rates of the discussed radiative transitions:
\be
f(W_{b0} \gamma):f(W_{b0}' \gamma): f(W_{b1} \gamma):f(W_{b2} \gamma)= {3 \over 4} \, \omega_0^3 : {1 \over 4} \, \omega_2^3 : 3 \, \omega_1^3 : 5 \omega_2^3~,
\label{fgr}
\ee
where $\omega_{0,1,2}$ are the photon energies in the corresponding transitions:
$\omega_0 \approx 305\,MeV$, $\omega_1 \approx 260\,MeV$, and $\omega_2 \approx 215\,MeV$. If the kinematical $\omega^3$ factors are taken into account in Eq.(\ref{fgr}), then the ratio of the rates is estimated approximately as 8.5 : 1 : 21 : 20, rather than as 3:1:12:20 in the case where one ignores these factors on the grounds that their difference is just another effect of the heavy quark spin interaction. This in fact illustrates the range of uncertainty in the predictions based on the heavy quark limit for the discussed transitions. However in spite of such an uncertainty, the presented estimates clearly indicate the relative feasibility of observing the yet hypothetical $W_{bJ}$ resonances.

It can be noted that the isoscalar counterparts of the $W_{bJ}$ resonances can in principle be also sought for in the radiative transitions from $\Upsilon(5S)$. Such C-even molecular resonances $X_{bJ}$ were considered in Ref.~\cite{bgmmv}. One can expect however that the rates for the radiative transitions are small as compared to those for the isovectors. Indeed, the $b$ quark and the antiquark are slow in both the initial and the final state in the transition. Thus an emission of the photon by the heavy quarks can be neglected, and the photon is radiated by the current of the light $u$ and $d$ quarks. In the latter current the isoscalar part is only 1/3 of the isovector one in the amplitude, and one estimates
\be
{\Gamma[\Upsilon(5S) \to X_{bJ} \, \gamma] \over \Gamma[\Upsilon(5S) \to W_{bJ} \, \gamma]} \approx 1/9~.
\label{ratgam}
\ee
However, unlike the $W_{bJ}$ states, the isoscalar resonances $X_{bJ}$ can contain an admixture of $\chi_{bJ}$ bottomonium states, and thus can be possibly be produced with an observable rate 
 in high-energy $p \bar p$ and $p p$ collisions at the Tevatron and LHC~\cite{bgmmv}, which provides a viable option for their discovery.
 
In summary. The relations between the rates of radiative transitions from $\Upsilon(5S)$ to the (yet hypothetical) isovector molecual bottomonium resonances $W_{bJ}$ with $G=-1$ are considered in the HQSS limit using the decomposition of the states in terms of the heavy and SLB spin structure. It is argued that the $\Upsilon(5S)$ resonance is dominantly a $1^-_H \otimes 0^+_{SLB}$ state and that the existing data do not contradict such an assignment. The relations between the rates of the radiative transitions are then found by applying the HQSS to the radiative processes.

I am grateful to Alexander Bondar for many enlightening discussions. This work is supported, in part, by the DOE grant DE-FG02-94ER40823.

\end{document}